\documentclass[sigconf]{acmart}

\usepackage{type1cm}     
\usepackage{graphicx}     
\usepackage{xspace}     
\usepackage{booktabs}     
\usepackage{multi row}     
\usepackage[font={bf}, tableposition=top]{caption}     
\usepackage{bold-extra}     
\usepackage{siunitx}          
\usepackage[vlined,linesnumbered,ruled,noend]{algorithm2e}     
\usepackage{microtype}    
\usepackage{units}     
\usepackage{mathtools}     
\usepackage{amssymb}     
\usepackage[show]{chato-notes}
\usepackage{relsize}
\usepackage{graphicx}
\usepackage{caption}
\usepackage{booktabs}     
\usepackage{multirow}
\usepackage{mathtools} 
\usepackage[]{inputenc}
\usepackage{amssymb}
\usepackage{amsmath}
\usepackage{siunitx} 
\usepackage[]{algorithm2e}
\usepackage[font={small}]{subfig} 
\usepackage{xfrac}

\newcommand{\spara}[1]{\smallskip\noindent\textbf{#1}}

\newenvironment {squishlist}
{\begin{list}{$\bullet$}
  { \setlength{\itemsep}{0pt}
     \setlength{\parsep}{3pt}
     \setlength{\topsep}{3pt}
     \setlength{\partopsep}{0pt}
     \setlength{\leftmargin}{1.5em}
     \setlength{\labelwidth}{1em}
     \setlength{\labelsep}{0.5em} } }
{\end{list}}

\acmConference[WebSci\textquotesingle17]{9th International ACM Web Science Conference}{June
 26--28 2017}{Troy, NUY, USA}

\setcopyright{rightsretained}

\begin{document}

\copyrightyear{2017} 
\acmYear{2017} 
\setcopyright{acmlicensed}
\acmConference{WebSci '17}{June 25-28, 2017}{Troy, NY, USA}\acmPrice{15.00}\acmDOI{http://dx.doi.org/10.1145/3091478.3091515}
\acmISBN{978-1-4503-4896-6/17/06}

\title{Factors in Recommending Contrarian Content\\on Social Media}

\author{Kiran Garimella}
\affiliation{%
  \institution{Aalto University}
  \city{Helsinki} 
  \country{Finland} 
}
\email{kiran.garimella@aalto.fi}

\author{Gianmarco De~Francisci~Morales}
\affiliation{%
  \institution{Qatar Computing Research Institute}
  \city{Doha} 
  \country{Qatar} 
}
\email{gdfm@acm.org}

\author{Aristides Gionis}
\affiliation{%
  \institution{Aalto University}
  \city{Helsinki} 
  \country{Finland} 
}
\email{aristides.gionis@aalto.fi}

\author{Michael Mathioudakis}
\affiliation{%
  \institution{Aalto University}
  \city{Helsinki} 
  \country{Finland} 
}
\email{michael.mathioudakis@aalto.fi}

\renewcommand{\shortauthors}{Garimella, De Francisci Morales, Gionis, Mathioudakis}

\begin{abstract}
Polarization is a troubling phenomenon that can lead to societal divisions and hurt the democratic process.
It is therefore important to develop methods to reduce it.

We propose an algorithmic solution to the problem of reducing polarization. 
The core idea is to expose users to content that challenges their point of view, 
with the hope broadening their perspective, and thus reduce their polarity.
Our method takes into account several aspects of the problem,
such as the estimated polarity of the user,
the probability of accepting the recommendation, 
the polarity of the content,
and popularity of the content being recommended.

We evaluate our recommendations via a large-scale user study on Twitter users
that were actively involved in the discussion of the US elections results.
Results shows that, in most cases, the factors taken into account in the recommendation
affect the users as expected, and thus capture the essential features of the problem.
\end{abstract}

\maketitle

\section{Introduction}
\label{sec:intro}

Polarization around controversial issues is a well-studied phenomenon 
in the social sciences~\citep{isenberg1986group}.
Social media have arguably amplified polarization, 
thanks to the scale of discussions and their publicity~\cite{garimella2017long}. 
This paper studies how to reduce polarization 
on social media by recommending \emph{contrarian} content, i.e., content that expresses 
a point-of-view opposing the one held by the target user.
In particular, we examine which features might be used to develop such a content recommender system.

We focus on controversial issues that create discussions online.
Usually, these discussions involve a fair share of ``retweeting'' or ``sharing'' opinions of authoritative figures with whom the user agrees.
Therefore, it is natural to model the discussion as an \emph{endorsement graph}:
a vertex $u$ represents a user, and 
a directed edge $(u,v)$ represents the fact that user~$u$ endorses the opinion of user $v$.

Due to phenomena such as 
homophily, confirmation bias, and selective exposure,
social media often create 
echo chambers~\citep{garrett2009echo, ebbandflow2017}.
These chambers, in turn, 
cultivate isolation and misunderstanding in society~\cite{sunstein2009republic},
and deepen its polarization.


A potential solution to this problem
is to encourage users to consider
points of view different from their own.
Thus, in this paper, 
we study methods to recommend content items 
(e.g., news articles, opinion pieces, blog posts) 
that express a contrarian point of view, while at the same time being appealing to the target user.


In particular, given metrics that measure the polarization of users and items
(such as those proposed in recent research~\citep{garimella2016quantifying}),
our goal is to recommend an item that nudges the user towards the opposite polarity.
That is, we seek to propose content produced by a user $v$ to another user $u$, 
thus informing $u$ of a different viewpoint, and hoping that $u$ will endorse $v$.

Clearly, some content is more likely to be endorsed than other.
For instance, people in the ``center'' might be easier to convince than people 
on the two extreme ends of the political spectrum~\cite{liao2014can}.
We take this issue into account by modeling the \emph{acceptance probability} 
for a recommendation as a separate component of the model.

We blend these factors, together with other signals such as topic and popularity, to create a ranked list of recommendations.
Our solution employs a well-known weighted rank-aggregation algorithm at its core~\citep{pihur2009rankaggreg}.

We evaluate our proposal by running an online user study with Twitter users.
We focus on the recent 2016 US presidential elections, and 
generate recommendations for the thousands of users involved in this highly-polarizing controversial discussion.
The results of the study show that the two main factors used in the recommendation, the polarity and the acceptance probability models, are predictive of the responses of the users.

In summary, we make the following contributions:
\begin{squishlist}
\item We study the problem of bridging echo chambers algorithmically, in a language- and domain-agnostic way.
Previous studies that address this problem focus mostly on understanding \emph{how} 
to recommend content to an ideologically opposite side, 
while we focus on \emph{which} contrarian content to recommend.
We believe that the two approaches complement each other 
in bringing us closer to bursting filter bubbles.
\item We build on top of results from recent user 
studies~\cite{munson2013encouraging,liao2014expert,vydiswaran2015overcoming} 
on how users prefer to consume content from opposing views, and 
formulate the task as 
a content-recommendation problem based on an endorsement graph, 
while also taking into account the acceptance probability of a recommendation.
\item We evaluate the proposed solution via a user study on Twitter users, and 
demonstrate the validity of the main factors involved in the recommendation.
\end{squishlist}


\section{Related work}
Although the Web was envisioned as a place 
of open discussions on a wide range of topics, 
many people tend to restrict themselves to viewing and sharing information that conforms with their beliefs.
A wide body of recent studies has 
explored~\cite{adamic2005political,conover2011political} and 
quantified~\cite{garimella2016quantifying} 
the notions of ``filter bubble'' and ``echo chambers''.
%

\citet{munson2013encouraging} created a browser widget that measures 
the bias of users based on the news articles they read.
Their study shows that users are willing to slightly change views 
once they are shown their biases.
\citet{graells2013data} show that mere display of contrarian content has negative emotional effect.
To overcome this effect, they propose a visual interface for making
recommendations from a diverse pool of users, 
where diversity is with respect to user stances on a topic.
\citet{graells2014people} propose to find topics that may be of interest to both sides 
by constructing a topic graph. They define intermediary topics to be those topics 
that have high betweenness centrality and topic diversity.
\citet{park2009newscube} propose methods for presenting multiple aspects of news to reduce bias.

Most relevant to this work is the recent 
study about the problem of reducing the overall polarization of a controversial topic 
in a network~\cite{garimella2017reducing}.
The study tries to find the best edges that can be added to an endorsement graph so that the polarization score of the network is reduced.
In this paper, we focus on reducing the polarization of an individual user (local objective), instead of the entire network (global objective).

There have also been a number of demos and systems:
Wall Street Journal's \emph{Blue feed-Red feed}\footnote{\footnotesize\url{http://graphics.wsj.com/blue-feed-red-feed/}} raises awareness about the extent to which viewpoints on a matter can differ, by showing side-by-side articles expressing very liberal and very conservative viewpoints;
\emph{Politecho}\footnote{\url{http://politecho.org/}} displays how polarizing the content on a user's news feed is when compared to their friends';
\emph{Escape your bubble}\footnote{\url{https://www.escapeyourbubble.com/}} 
is a browser extension to add hand-curated content from the opposite side in Facebook;
automated bots have been created to respond to posts containing harassment or fake news,\footnote{\url{http://wpo.st/4kVR2}, \url{https://goo.gl/Xl6x9t}} with an attempt to de-polarize the discussion and educate users.
Moreover, new social media platforms have been proposed that aim to be designed in such a way to encourage discussions and debates, such as 
the Filterburst project,\footnote{\url{http://www.filterburst.com/}}
Rbutr,\footnote{\url{http://rbutr.com/}} where users can post rebuttals of other urls, and 
a wikipedia for debates.\footnote{\url{http://www.debatepedia.org/en/index.php/Welcome_to_Debatepedia\%21}}

The proposed method differs from existing ones in many ways. 
First, our approach is completely algorithmic, 
unlike most demos listed above, which involve manual curation.
Second, as discussed above, 
it builds on top of existing research and incorporates key findings of previous work.

\section{Preliminaries}
A topic of discussion is identified as the set of tweets that satisfy a text query -- e.g., all tweets that contain a specific hashtag.
We represent a topic with an \emph{endorsement graph} $G(V, E)$, 
where vertices $V$ represent users
and edges $E$ represent \emph{endorsements}.

It has been shown that
an endorsement graph captures well the extent to which 
a topic is controversial~\citep{garimella2016quantifying}.
In particular, 
the endorsement graph of a controversial topic 
has a \emph{multimodal clustered structure}, where each cluster of vertices
represents one viewpoint on the topic.
As we focus on two-sided controversies, we identify the two sides of a controversial topic by employing a \emph{graph-partitioning} algorithm, 
which partitions the graph into \emph{two} subgraphs.
In this work, we specifically focus on recommending content in the form of news items, 
such as articles, blog posts, and opinion pieces.
The item pool for the recommendation comprises all the links shared by the active users 
during the observation window.

\spara{User polarization score.}
We use a recently-proposed metho\-do\-logy 
to define the polarization score for each user in the graph~\cite{garimella2016tscpaper}.
The score is based on the expected hitting time of a random walk that starts 
from the user under consideration and ends on a high-degree vertex on either side.
Typically, in a retweet graph, 
high-degree vertices on each side are indicators of authoritative content generators.
We denote the set of the $k$ highest degree vertices on each side by $X^+$ and $Y^+$.
Intuitively, a vertex is assigned a score of higher absolute value (closer to $+1$ or $-1$), 
if, compared to other vertices in the graph, 
it takes a very different time to reach a high-degree vertex on either side ($X^+$ or $Y^+$) (in terms of information flow). 
Specifically, for each vertex $u\in V$ in the graph, 
we consider a random walk that starts at $u$, and 
estimate the expected number of steps, $l_u^{_X}$ 
before the random walk reaches any high-degree vertex in $X^+$.
Considering the distribution of values of $l_u^{_X}$ across all vertices $u \in V$, 
we define $\rho^{_X}(u)$ as the fraction of vertices $v\in V$ with $l_v^{_X} < l_u^{_X}$.
We define $\rho^{_Y}(u)$ similarly.
Obviously, we have $\rho^{_X}(u), \rho^{_Y}(u) \in [0,1)$.
The polarization score of a user is then defined as
\begin{equation}
	\rho(u) = \rho^{_X}(u) - \rho^{_Y}(u) \quad \in (-1, 1) .
\end{equation}
Following this definition, a vertex that is close to high-degree vertices $X^+$, 
compared to most other vertices, will have $\rho^{_X}(u) \approx 1$;
on the other hand, if the same vertex is far from high-degree vertices $Y^+$, 
it will have $\rho^{_Y}(u) \approx 0$;  
leading to a polarization score $\rho(u) \approx 1 - 0 = 1$.
The opposite is true for vertices that are far from $X^+$ but close to $Y^+$; 
leading to a polarization score $\rho(u) \approx -1$.

\spara{Item polarization score.}
Once we have obtained polarization scores for users in the graph, it is straightforward to derive a similar score for content items shared by these users.
Specifically, we define the polarization score of an item $i$ as the average of the polarization scores of the set of users who have shared $i$, denoted by $U_i$:
\begin{equation}
	\rho(i) = \frac{1}{|U_i|} \displaystyle\sum_{u \in U_i} \rho(u)  \quad \in (-1, 1) .
\end{equation}

\spara{Acceptance probability.}
Not all recommendations are agreeable, 
especially if they do not conform to the user's beliefs.
To reduce these effects, we define an acceptance probability, 
which quantifies the degree to which a user is likely to endorse the recommended content.
We use the item and user polarization scores defined above to estimate the likelihood that a target user $u$ endorses (i.e., retweets) the recommended item $i$.
We build an acceptance model by adapting a similar one based 
on the feature of user polarization~\citep{garimella2017reducing}.
High absolute values of user polarization
(close to $-1$ or $+1$) indicate that the user belongs clearly to one side of the controversy, 
while middle-range values (close to $0$) indicate that the user is in the middle of the two sides.
It was shown that users from either side 
accept content from different sides with different probabilities, and 
 these probabilities can be inferred from the graph structure~\cite{garimella2017reducing}.
For example, a user with polarization close to $-1$ is more likely to endorse a user with a negative polarization than a user with polarization $+1$. 
This intuition directly translates to endorsing items, and therefore can be used for our recommendation problem.

Based on this intuition, we define the acceptance probability $p(u,i)$ 
of a user $u$ endorsing item $i$ as
\begin{equation}
p(u,i) = \sfrac{ N_{e}(\rho(u),\rho(i)) } { N_{x}(\rho(u),\rho(i)) } ,
\end{equation}
where $N_{e}(\rho(u),\rho(i))$ and $N_{x}(\rho(u),\rho(i))$ are the number of times a user with polarity $\rho(u)$ has endorsed or was exposed to (respectively) 
content of polarity $\rho(i)$. 
In practice, the polarity scores are bucketed to smooth the probabilities.


\section{Recommendation factors}

This section describes the factors used to generate recommendations.
Though our main focus is to connect users with content that expresses a contrarian point of view, we also want to maximize the chances of such a recommendation being endorsed by the user.
%
We take into account several factors: reduction in polarization score of the target user;
exclusivity of the candidate items (polarity of the items);
acceptance probability of recommendation based on polarization scores;
topic diversity;
popularity/quality of the candidate item.
Next, we describe these factors in more detail.

\spara{Reduction of user polarization score.}
The maximum reduction of user polarization score is achieved by putting the user in contact 
with an authoritative source from the opposing side.
Leveraging this idea, we build a list of items $L_1$
by considering items shared by high degree nodes on the opposite side of the target user, 
and ranking them by the potential decrease in user polarization score. 

\spara{Exclusivity on either side.}
We consider items that are almost exclusively shared by one of the sides.
Specifically, we denote by $n_i^X$ and $n_i^Y$ the number of users 
who shared each item $i$ on side $X$ and $Y$, respectively.
For each side, we generate a list $L_2$ ranked 
by the ratio of shares $\sfrac{n_i^X}{n_i^Y}$ (for side $X$) and $\sfrac{n_i^Y}{n_i^X}$ (for side $Y$).

\spara{Acceptance probability.}
For a given user $u$, all items sorted in decreasing order of acceptance probability $p(u,i)$ make up list $L_3$.

\spara{Topic diversity.}
We want to ensure that the recommendations are topically diverse.
To achieve this, for each user, we compute a vector $t_u$ 
that contains the topics extracted from the tweets written and the items shared by the user.
Similarly, we extract a vector of topics $t_i$ for each item. 
Topics are defined as {\em named entity}, 
and we extract them using the tool tagme.\footnote{\url{https://services.d4science.org/web/tagme}}
Given a user vector $t_u$, we compute the cosine similarity with all item vectors $t_i$, 
and rank items in increasing order of cosine similarity (list $L_4$).

\spara{Popularity on either side.}
Finally, we take into account the popularity of the recommended items, 
so that users receive content that is popular and, likely, of good quality.
For each item, we compute a popularity score as the maximum number of retweets 
obtained by a tweet that contains this item.
We produce list $L_5$ of items in decreasing popularity score.

\spara{Rank Aggregation.}
Given the 5 ranked lists discussed above,
we use a weighted rank-aggregation scheme to generate the final recommendations.
The intuition behind using rank aggregation is that items that are highly ranked in many lists, 
are also highly ranked in the output list.
In particular, we use a weighted rank-aggregation technique proposed by \citet{pihur2009rankaggreg},
whose goal is to minimize the objective function
\begin{equation}
	\phi(\delta) = \sum_{i=1}^{5} w_i d(\delta,L_i) ,
\end{equation}
where $\delta$ is the optimal ranked output list, $d$ is any distance function 
(we use the Spearman footrule distance), and 
$w_i$ are the importance weights of each list.
We can set the weights to generate highly contrarian recommendations 
(by giving large weights to $L_1$ and $L_2$) or 
recommendations that are likely to be accepted (by giving large weight to $L_3$).


\section{Evaluation}
\label{sec:evaluation}


\spara{Dataset.}
We collect all tweets containing the hashtag 
\texttt{\small\#USelections}, used in discussions about the US presidential elections during Nov 9--12, 2016.
From the \num{6.2}\,M tweets collected, 
we build an endorsement graph with \num{6764} nodes (users) and \num{9896} edges (retweets).
To filter out noise, the graph contains an edge between two users only if 
at least 
\num{5} retweets between the two users occur.
We partition the  graph to obtain the two sides by using 
METIS~\cite{karypis1995metis}.
For recommendation items (urls), we consider items 
that have been shared at least $5$ times in our dataset.
The final pool contains \num{10210} candidate items, which include news articles, blog posts, opinion pieces, etc. 


\begin{figure}[t]
\centering
\includegraphics[width=0.8\columnwidth]{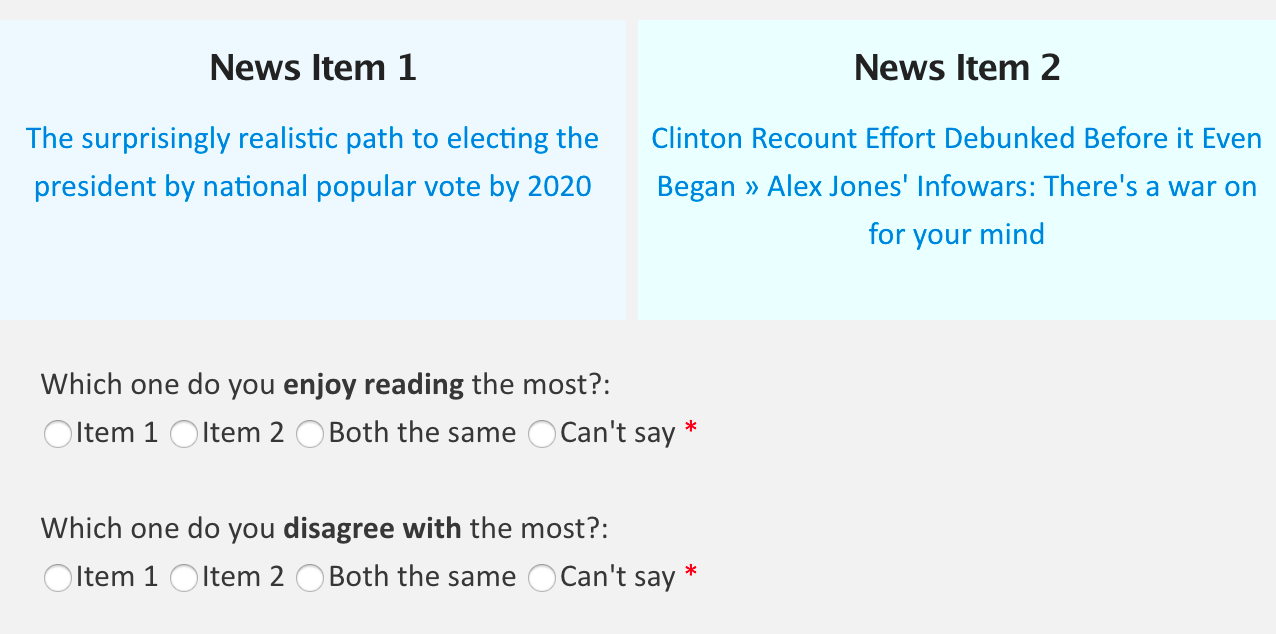}
\caption{Screenshot of the interface shown for a user with a high polarity on the political left (Democrat).}
\label{fig:screenshot_dem}
\vspace{-\baselineskip}
\end{figure}


\begin{figure}[t]
\centering
\includegraphics[width=0.8\columnwidth]{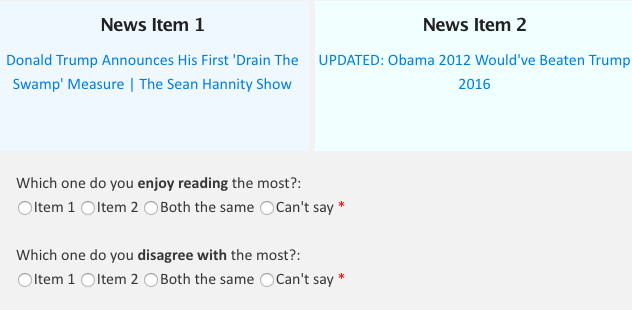}
\caption{Screenshot of the interface shown for a user with a high polarity on the political right (Republican).}
\label{fig:screenshot_rep}
\vspace{-\baselineskip}
\end{figure}

\spara{User study.}
We run an online user study involving all \num{6764} users in the dataset with the aim of evaluating 
how users percieve the two main conflicting factors proposed, i.e. the contrarian features ($L_1$, $L_2$) and acceptance features ($L_3$).
For each user in the study, we generate two recommended items that are personalized based on their Twitter activity:
one item is highly contrarian, 
while the other is more likely to be accepted, according to our model.
%
In more detail, by using the methodology described above, we compute two recommendations for each user: in the first one we give a high weight ($60\%$) to contrarian features ($L_1$ and $L_2$), while in the second one we give high weight (60\%) to acceptance probability ($L_3$). We distribute the remaining $40\%$ equally among other features.

The main research questions we investigate are:
($i$) is a high acceptance probability factor predictive of content with higher acceptance?
and ($ii$) are contrarian factors predictive of more disagreement with the user?
To simplify the task for the user, we set up the user study as a relative comparison between the two recommendations, rather than asking for absolute judgments.
Since the two recommendations are generated completely independently, we assume that they do not influence the users decision making process in choosing one over the other.

We create a web form\footnote{\url{http://bit.ly/2jOQBxP}} with two recommended items, customized for each user, with the item weighted by the acceptance features shown on the left and contrarian features on the right.
Figures~\ref{fig:screenshot_dem} and~\ref{fig:screenshot_rep} show two instances of the web form.
Looking at Figure~\ref{fig:screenshot_dem}, given the left-leaning political affiliation of the user, 
the recommendation on the left side (News item 1) 
looks more agreeable than the recommendation on the right side (News item~2).
The opposite is true for Figure~\ref{fig:screenshot_rep}, which targets a right-leaning user.

We contacted users on Twitter with the following private message: 
``@username We are scientists studying social media. Would u like to help science by participating in a survey? \nolinkurl{http://bit.ly/XXXXX'}',
and waited for two weeks for them to respond.
In total, we sent around \num{6700} messages and received \num{93} valid responses after removing duplicates ($1.4\%$ response rate).

\begin{table}[t]
\centering
\small
\caption{Results from the user study.}
\label{tab:results}
\begin{tabular}{l c c c c}
\toprule
Main factor & Item1 & Item2 & Both & Can't \\
	& (Acceptance) & (Contrarian) & the same & say \\
\midrule
Enjoy & 51    & 19       & 8          & 15       \\
Disagree & 22    & 57       & 7          & 7       \\
\bottomrule
\end{tabular}
\vspace{-\baselineskip}
\end{table}

Our expectation is that users enjoy reading the item with high acceptance probability, and disagree with the contrarian item.
The results, summarized in Table~\ref{tab:results}, confirm our expectations.
Indeed, most users enjoy reading the item with high acceptance, and disagree with the contrarian item.
Specifically, $44$ out of the $93$ users ($47\%$) reported that at the same time they enjoy the first item, and disagree with the second.
For a few users (n=7), we were able to generate enjoyable recommendations that they disagreed with.
While this was not the goal of the specific user study, it is indeed our ultimate goal, and thus these results are highly encouraging.

\spara{Acknowledgements.}
This work has been supported by the Academy of Finland project ``Nestor'' (286211) and the EC H2020 RIA project ``SoBigData'' (654024).

\vspace{-3mm}
\bibliographystyle{ACM-Reference-Format}
\bibliography{biblio}

\end{document}